\RequirePackage{lineno}
\documentclass[aps,floatfix,prc,twocolumn,showpacs,showkeys,amsmath,amssymb,superscriptaddress]{revtex4}

\usepackage{soul}
\usepackage{graphicx}	
\usepackage{xspace}     
\usepackage{amsmath}
\usepackage{hyperref}
\usepackage{color}
\usepackage{multirow}
\usepackage{booktabs,array}

\newcolumntype{C}[1]{>{\centering\let\newline\\\arraybackslash\hspace{0pt}}m{#1}}

\newcommand{\dhcs}{dihadron correlations\xspace}
\newcommand{\Dhcs}{Dihadron correlations\xspace}

\newcommand{\dhc}{dihadron correlation\xspace}

\newcommand{\jlcs}{jet-like correlations\xspace}

\newcommand{\dphi}{$\Delta\phi$\xspace}
\newcommand{\deta}{$\Delta\eta$\xspace}

\newcommand{\vn}{$v_{n}$\xspace}
\newcommand{\vnum}[1]{$v_{#1}$\xspace}

\newcommand{\sref}[1]{section~\ref{#1}}
\newcommand{\fref}[1]{figure~\ref{#1}}
\newcommand{\tref}[1]{table~\ref{#1}}

\newcommand{\Fref}[1]{Figure~\ref{#1}}

\newcommand{\pp}{$p$+$p$\xspace}

\newcommand{\Au}{Au+Au\xspace}

\newcommand{\dAu}{$d$+Au\xspace}

\newcommand{\sNN}{$\sqrt{s_{\mathrm{NN}}}$\xspace}

\newcommand{\ns}{near-side\xspace}
\newcommand{\as}{away-side\xspace}

\newcommand{\pttrigrange}[2]{#1 $< p_T^{\mathrm{t}} <$ #2~GeV/$c$\xspace}
\newcommand{\ptassocrange}[2]{#1 $< p_T^{\mathrm{a}} < $ #2~GeV/$c$\xspace}

\newcommand{\etarange}[1]{$|\eta|<$#1\xspace}

\begin{document}

\title{Reexamining the iconic dihadron correlation measurement demonstrating jet quenching}

\author{Christine Nattrass} 
\affiliation{
University of Tennessee, Knoxville, Tennessee 37996, USA. \\
}
\date{\today}

\begin{abstract} 
Early measurements at the Relativistic Heavy Ion Collider (RHIC) demonstrated jet quenching through the suppression of pairs of high momentum hadrons.  These \dhcs have a large correlated background. As understanding of the background improved, it was recognized in the field that a significant term was omitted from the background and several \dhc results were quantitatively and qualitatively incorrect.  The original measurements demonstrating jet quenching have not been revisited.  These measurements are repeated in this paper in a kinematic range similar to the original measurement using publicly available data, applying current knowledge about the background.  The new results are qualitatively consistent with the previous results, demonstrating complete suppression of the \as within uncertainties.

\end{abstract}

\maketitle

\section{Introduction}

The Quark Gluon Plasma (QGP), a hot, dense liquid of quarks and gluons, is formed in high energy heavy ion collisions~\cite{Adcox:2004mh,Adams:2005dq,Back:2004je,Arsene:2004fa}.  One of the main signatures of the QGP is jet quenching, partonic energy loss in the medium through collisions with medium partons and gluon bremsstrahlung.  This leads to a suppression of jet fragments carrying a large fraction of the parent parton's momentum (high $z = p_{T}^{hadron}/E_{jet}$) and an enhancement of jet fragments carrying a small fraction of the parent parton's momentum (low $z$).  There is extensive experimental evidence for jet quenching, including the suppression of high momentum hadrons relative to expectations from proton-proton collisions~\cite{Adams:2003kv,Adler:2003qi,Back:2004bq,Aamodt:2010jd,CMS:2012aa,Khachatryan:2016odn}, the suppression of hadrons $180^{\circ}$ away in azimuth from a high momentum hadron~\cite{Adler:2002tq,Adams:2003im,Adams:2004wz}, and the direct observation of an asymmetry in the energy of di-jet pairs~\cite{Aad:2010bu,Chatrchyan:2011sx}.

The role of high momentum \dhcs in providing experimental evidence for jet quenching is difficult to overstate.  Over forty measurements to date of correlations of hadrons, photons, and leptons with high momentum hadrons in nucleus-nucleus collisions have been published by experiments at RHIC and the LHC.  The paper reporting the suppression of particles $180^{\circ}$ away from a high momentum particle has over 750 citations~\cite{Adams:2003im}.  This paper continues to be cited as evidence for jet quenching and its iconic plot is shown, both among those studying the QGP and outside the field.  This is despite widespread knowledge within the community that the background subtraction omitted a key term, the third order coefficient of the Fourier distribution of azimuthal anisotropy, \vnum{3}~\cite{Sorensen:2010zq,Alver:2010gr}.  While there is substantial experimental evidence establishing jet quenching~\cite{Connors:2017ptx}, there has been no reanalysis in a similar kinematic regime to the initial paper to determine if evidence for jet suppression is robust at the momenta in the original study.

Data in a similar kinematic regime are used in this paper in order to update the measurement in~\cite{Adams:2003im} with current knowledge about the background in \dhcs.  The form of the background is briefly reviewed and publicly available data~\cite{Agakishiev:2010ur,Abelev:2009af} are used to produce updated \dhcs in a similar kinematic regime to~\cite{Adams:2003im}.  This updated plot is qualitatively consistent with the original.  There are some limitations in the analysis possible with publicly available data because the first order coefficient of the azimuthal anisotropy, \vnum{1}, is poorly constrained.  Two approaches to the background subtraction are used, one which includes \vnum{1} in a fit to the background-dominated region but is susceptible to unstable fits and one approach which uses independent measurements of the \vn and fixes $v_1 = 0$.  

\section{Separating signal and background in \dhcs}\label{Sec:method}
In high momentum \dhcs, a high momentum trigger particle is selected and the distribution of particles relative to that trigger particle in azimuth, $\Delta\phi = \phi^{trigger} - \phi^{associated}$, is measured.  The correlation function is dominated by $2\to 2$ processes.  The signal from \jlcs is typically described as a \ns peak ($\Delta\phi<\pi/2$) from particles produced by fragmentation of the same jet as the trigger particle and an \as peak ($|\Delta\phi - \pi| <\pi/2$) from its partner jet.

The correlation function will have contributions from all physical correlations in the event.  Hanbury-Brown-Twiss (HBT) correlations, quantum correlations between particles from the same source, are suppressed by a difference between the momentum of the trigger and associated particles~\cite{Lisa:2005dd,Lisa:2008gf}.  Correlations between decay daughters and electron-positron pairs from conversions are suppressed by focusing on high momentum.  Any remaining contributions from HBT and decays are effectively considered part of the signal.  In heavy ion collisions, both the jet signal and the flow-modulated combinatorial background are correlated with the reaction plane, the former due to the path length dependence of partonic energy loss and the latter due to hydrodynamical flow.  Spatial asymmetries in the initial overlap region between two nuclei are converted to momentum anisotropies in the final state~\cite{Voloshin:2008dg}.  Since both the signal and the background are correlated with the reaction plane, they will be correlated with each other.  Furthermore, the minimal threshold on the momentum of the trigger particle increases the probability that it was produced in a jet but does not guarantee it.  There is therefore a correlated background which can be described by its Fourier decomposition~\cite{Bielcikova:2003ku}
\begin{equation}
 B(\Delta\phi) = B_0 \Bigg(1+
 2\sum_{n=1}^{\infty} v_{n}^{a} {v}_{n}^{t} \cos(n \Delta\phi) 
\Bigg).
\end{equation}
\noindent
In the case that the correlated background is dominated by flow, the \vn in these correlations are the same as independent measurements of the \vn due to flow.  Initially, the odd $n$ terms were assumed to be zero because the average distribution of nucleons in nucleus is symmetric.  It was later proposed that fluctuations in the positions of the nucleons could lead to odd $n$ \vn~\cite{Alver:2010gr,Sorensen:2010zq}, which were later observed~\cite{Adare:2011tg,ALICE:2011ab}.

The omission of \vnum{3} in particular led to two artifacts, the ``ridge'' on the \ns~\cite{Abelev:2009af,Alver:2009id}, a structure which was correlated in \dphi but roughly independent of \deta, and the ``Shoulder'' or ``Mach cone'' on the \as~\cite{Abelev:2009af,Adare:2007vu,Adare:2008ae,Afanasiev:2007wi,Agakishiev:2010ur}, a dip at $\Delta\phi \approx \pi$ with two peaks additional peaks offset from $\Delta\phi \approx \pi$.  These effects were considerable for trigger momenta \pttrigrange{4}{6} and associated momenta \ptassocrange{2}{4}, the region used for the original \dhc demonstrating jet quenching~\cite{Adams:2003im}, motivating a reconsideration of the signal in this range.

Background subtraction has generally been done using the assumption that the yield is zero near $\Delta\phi\approx 1 $ combined with \vn from independent measurements, called the Zero-Yield-At-Minimum (ZYAM) method~\cite{Adams:2005ph}.  There have since been method developments to avoid these assumptions~\cite{Sickles:2009ka,Sharma:2015qra}.

The correlation function in \Au collisions at \sNN = 200 GeV in~\cite{Adams:2003im} is for \pttrigrange{4}{6} and \ptassocrange{2}{4} and pseudorapidities \etarange{0.7}.  The precise centrality range is not given, however, other STAR papers using the same data set use 0--5\% central collisions~\cite{Adams:2003kv} and the thesis including these measurements includes correlations from both 0--5\% and 0--10\% central collisions~\cite{phdthesismiller}.  The data from~\cite{Adams:2003im} are not publicly available before background subtraction.

The publicly available data set with a kinematic range similar to the original paper was chosen as the focal point of this analysis.  Correlation functions in 0--12\% central \Au collisions at \sNN = 200 GeV for \pttrigrange{4}{6}, \ptassocrange{2}{4}, and \etarange{1.0} from~\cite{Abelev:2009af} were measured as a function of both azimuth and pseudorapidity.  A correction for the pair acceptance in \deta was applied in~\cite{Abelev:2009af} which was not applied in~\cite{Adams:2003im}.  It would in principle be possible to undo this correction assuming a form of $a(\Delta\eta) = |2.0 - \Delta\eta/2.0|$.  Undoing this correction would increase the \as by nearly a factor of two, since it is nearly independent of \deta, and lead to a slight increase in the \ns.  Since the \as is observed to be consistent with zero, the correction is not undone in order to minimize manipulations of the data.  

A rapidity-even \vnum{1} due to flow has been observed to be comparable to \vnum{2} and \vnum{3}~\cite{Luzum:2010fb,ATLAS:2012at}.
The value of \vnum{1} is difficult to determine because there is no clear technique which can be used to separate contributions from flow and jets.  In~\cite{Luzum:2010fb,ATLAS:2012at} \dhcs are used where the trigger and associated particles are separated in \deta and assuming that the coefficient of $\cos(\Delta\phi)$ is $v_1^a v_1^t$.  The \ns is suppressed by the large \deta gap between the trigger and associated particles, but there is a residual contribution from the \as.  Any residual contribution from jets on the \as will lead to an artificially high \vnum{1}, particularly for the values in~\cite{Luzum:2010fb} which use particles as close as $\Delta\eta = 0.7$.  The data in~\cite{Luzum:2010fb} are from 20--60\% central \Au collisions and measurements in~\cite{ATLAS:2012at} have a somewhat larger \deta gap, but are at a different collision energy.  There therefore are no measurements which can be used to fix \vnum{1} for central \Au collisions at \sNN = 200 GeV.

The Near-Side Fit (NSF) method fits the background-dominated correlation function on the \ns at large \deta to determine the $B$ and \vn simultaneously~\cite{Sharma:2015qra} and \vnum{1} can be included.  The NSF method is applied to the data from \Au collisions from~\cite{Abelev:2009af} to determine the background, allowing a non-zero \vnum{1}.
The NSF method can be sensitive to the fit region.  
The ZYAM method is therefore also applied to the data in~\cite{Abelev:2009af} as a cross check using the \vnum{2} from~\cite{Abelev:2009af}. The \vnum{3} from 0--10\% central \Au collisions were estimated based on~\cite{Adare:2011tg}, increasing the uncertainty slightly to take the different momentum and centrality regions into account.  The \vnum{1} are assumed to be zero for the ZYAM method.

 The \vn from the fit and the ZYAM method are shown in \tref{Tab:vn}.  
 The value of \vnum{1} can be estimated from~\cite{Luzum:2010fb}, leading to approximately $v_1^a v_1^t = 0.008 \pm 0.004$.  This is substantially lower than that derived from the NSF method, although that may be due to the difference in centrality.  The uncertainties are estimated conservatively due to the residual contamination from the \as.  The standard deviation between values of $v_1^a v_1^t$ from the fit and from independent measurements are also shown in \tref{Tab:vn}.  The \vnum{1} is somewhat lower than the values from~\cite{Luzum:2010fb} while the \vnum{2} and \vnum{3} are somewhat higher.

\begin{table*}
\centering
\resizebox{\textwidth}{!}{\begin{tabular}{c || c | c | c || c | c | c || c | c | c}
  Method & $v_1^a v_1^t$ & $v_2^a$ from~\cite{Abelev:2009af} & $ v_2^t$ from~\cite{Abelev:2009af} & $v_2^a v_2^t$  & $v_3^a$ from~\cite{Adare:2011tg} & $ v_3^t$ from~\cite{Adare:2011tg} & $v_3^a v_3^t$ \\ \hline
NSF & $0.00225 \pm 0.00034$ & -- & -- & $0.00845 \pm 0.00033$  & -- &  -- & $0.00573 \pm 0.00033$ \\
$\sigma$ & -1.4 & -- & -- & +1.2  & -- &  -- & $^{+0.69}_{+0.61}$ \\ \hline
ZYAM& 0 &  $0.082 \pm 0.002$ & $0.077 \pm 0.019$ & $0.0063 \pm 0.0017$ & $0.067 \pm 0.010$ & $0.072 \pm 0.010$ & $0.0048 ^{+0.0015}_{-0.0013}$ \\
\end{tabular}}
\caption{\vn from the NSF method, number of standard deviations away from expectations, and \vn used in the ZYAM method.} 
\label{Tab:vn}
\end{table*} 

Unfortunately the \dAu data from~\cite{Abelev:2009af} are not available. Instead correlation functions from minimum bias \dAu collisions at \sNN = 200 GeV and 20--60\% \Au collisions at \sNN = 200 GeV for tracks with $\eta<1$ in two regions in \deta ($|\Delta\eta|<$~0.7 and 0.7~$<|\Delta\eta|<$~2.0) from~\cite{Agakishiev:2010ur} are used.  Data from \Au collisions are in bins of the angle between the trigger particle and the reconstructed reaction plane, $\phi_s = \phi^t - \psi$ and were reanalyzed in~\cite{Nattrass:2016cln} using the Reaction Plane Fit method~\cite{Sharma:2015qra}.  The data from \Au collisions from all bins relative to the reaction plane and \ptassocrange{2}{3} and \ptassocrange{3}{4} are combined to get the same momentum region as~\cite{Adams:2003im}.  A constant background is assumed for data from \dAu collisions.  Both the \dAu and the 20--60\% \Au collisions from~\cite{Abelev:2009af} are from $|\Delta\eta|<$~0.7.  The slightly different acceptance range could decrease the \ns yield, though such effects are likely negligible.  The correction for acceptance in \deta was not applied to these data.

\section{Results}

\begin{figure}
\begin{center}
\rotatebox{0}{\resizebox{8.0cm}{!}{
	\includegraphics{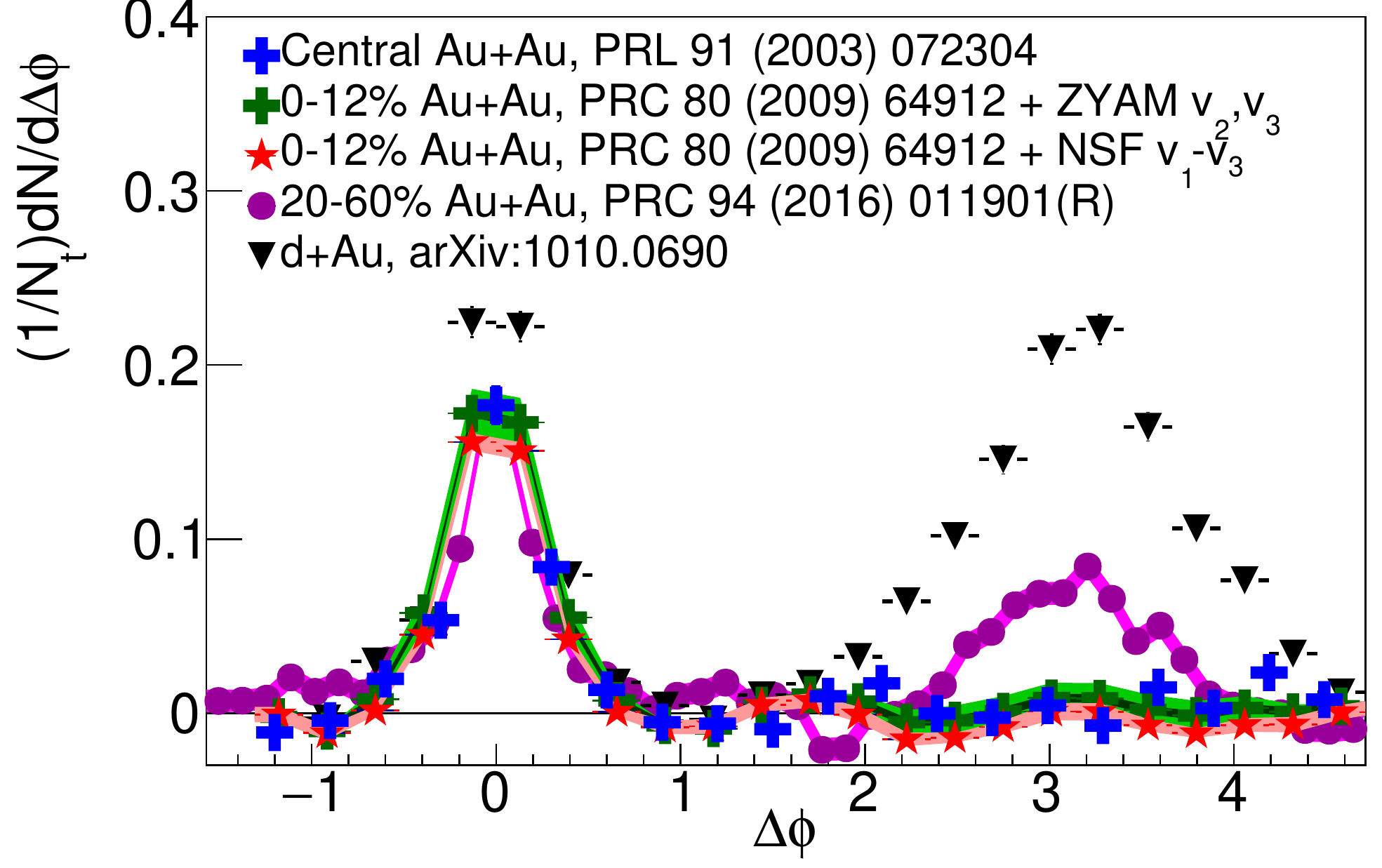}
}}
\caption{
\Dhcs with \pttrigrange{4}{6} and \ptassocrange{2}{4} from minimum bias \dAu collisions at \sNN = 200 GeV from~\cite{Agakishiev:2010ur}, 20--60\% central \Au collisions at \sNN = 200 GeV from~\cite{Agakishiev:2010ur} reanalyzed in~\cite{Nattrass:2016cln}, central \Au collisions at \sNN = 200 GeV from~\cite{Adams:2003im}, and 0--12\% collisions from~\cite{Abelev:2009af} reanalyzed using both the NSF method described in~\cite{Nattrass:2016cln} and the ZYAM method using $v_2$ from~\cite{Abelev:2009af} and $v_3$ from~\cite{Adare:2011tg}.
}
\label{Fig:NewIconicPlot}
\end{center}
\end{figure}

\Fref{Fig:NewIconicPlot} shows the background subtracted correlations comparable to the analysis in~\cite{Adams:2003im} but incorporating current knowledge about the background.  The slightly different acceptances in \deta described in \sref{Sec:method} could overestimate the \as in 0--12\% central \Au collisions by up to a factor of two relative to 20--60\% \Au collisions and \dAu collisions, however, it is consistent with zero and the previous results for both the ZYAM and NSF methods.  The slight differences in the \ns between minimum bias \dAu, 20-60\% central \Au, and 0--12\% central \Au collisions may be due to the differences in \deta and $\eta$ acceptance described in \sref{Sec:method}, or may be due to slight modifications of the \ns.  The largest difference is seen for the 20--60\% central \Au collisions, where there are no apparent shape modifications.  
For central \Au collisions, despite qualitatively different observations on the \as when taking \vnum{3} into account for slightly lower momentum particles, results on the \as in this particular kinematic regime happen to be  qualitatively consistent with those reported in~\cite{Adams:2003im}.

Some caution is warranted due to uncertainty in the appropriate value of \vnum{1}.  For both the NSF and ZYAM method shown above, the results are consistent.  The values of \vnum{1} in~\cite{Luzum:2010fb} would lead to a non-zero \as, but these values are probably overestimates due to contaimination of the \vnum{1} measurements by the jet signal and the difference in centrality.

There is not currently a motivation for emphasizing this kinematic regime.  The initial studies were limited in statistics, which motivated studying lower momentum trigger particles, where the probability that the trigger is from a jet is somewhat lower.  Contemporary data sets allow much higher momentum trigger particles.  Nonetheless, studies of low momenta associated particles still require precision subtraction of the large combinatorial background.  Ultimately this requires a better understanding of the contribution of \vnum{1}.  The reaction plane fit method~\cite{Sharma:2015qra} may be one way to overcome this, with more stability in the fit.

\begin{figure}
\begin{center}
\rotatebox{0}{\resizebox{8.0cm}{!}{
	\includegraphics{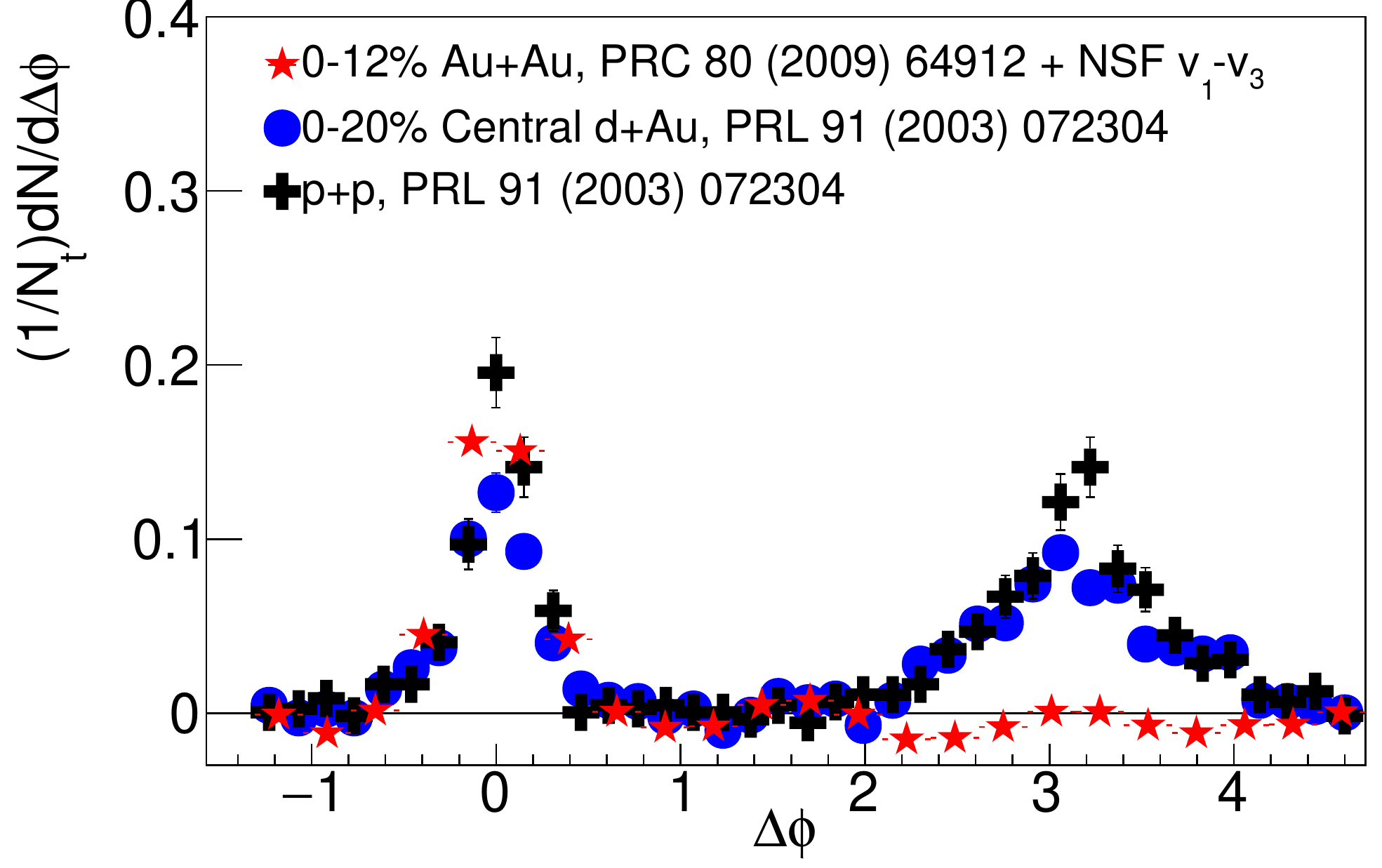}
}}
\caption{
\Dhcs with \pttrigrange{4}{6} and \ptassocrange{2}{4} from 0--20\% central \dAu collisions at \sNN = 200 GeV from~\cite{Adams:2003im}, \pp collisions at \sNN = 200 GeV from~\cite{Adams:2003im}, and 0--12\% collisions from~\cite{Abelev:2009af} reanalyzed using the NSF method described in~\cite{Nattrass:2016cln}.
}
\label{Fig:NewIconicPlot2}
\end{center}
\end{figure}

\Fref{Fig:NewIconicPlot2} compares the original data from \pp and \dAu collisions from~\cite{Adams:2003im} to the 0--12\% central \Au data~\cite{Abelev:2009af} from analyzed using the NSF method~\cite{Nattrass:2016cln}.  The NSF method is chosen because this approach takes \vnum{1} into account.  The slight difference in pseudorapidity acceptance between the data in~\cite{Adams:2003im} ($|\eta|<0.7$) and in~\cite{Agakishiev:2010ur} ($|\eta|<1.0$)  does not lead to a noticeable difference on the \ns but leads to an \as which is roughly 30\% lower in \dAu.  The data in~\cite{Abelev:2009af} are both in a different pseudorapidity range ($|\eta|<1.0$) from~\cite{Adams:2003im} and have an acceptance correction applied in $\Delta\eta$, leading to small differences on the \ns.  Since the \as is nearly completely suppressed, this does not impact the comparison of the 0--12\% data from~\cite{Abelev:2009af} to the \dAu data from~\cite{Adams:2003im}.  The message of \fref{Fig:NewIconicPlot2} is in agreement with the message in~\cite{Adams:2003im}.

\section{Conclusions}
There is comprehensive experimental evidence for jet quenching, but an early, iconic result has not been reassessed previously, long after it was widely recognized in the field that the background subtraction was incomplete.  Using data in a similar kinematic regime, this measurement was repeated using the field's current understanding of the data.  
 These results could be improved further by using a background subtraction technique which can constrain \vnum{1} better.
Fortunately for the community, the same qualitative conclusion can be drawn -- that the \as is nearly completely suppressed.

\section{Acknowledgements}
I are grateful to Redmer Bertens and Gabor David and for useful comments on the manuscript, Jo\"rn Putschke, Fuqiang Wang, and the STAR collaboration for providing data, and Gabor David for convincing me that it would be useful to the community to write this.  This work was supported in part by funding from the Division of Nuclear Physics of the U.S. Department of Energy under Grant No. DE-FG02-96ER40982.

\end{document}